\documentclass[sigconf, nonacm, authorsversion]{acmart}
\usepackage{tabularx}
\usepackage{booktabs} 
\usepackage{graphicx}
\usepackage{flushend}
\usepackage{booktabs} 
\usepackage{tablefootnote} 
\usepackage{booktabs} 
\usepackage{subcaption} 
\usepackage{amssymb} 
\usepackage{xcolor, colortbl} 
\usepackage{array,multirow,graphicx} 
\usepackage{makecell} 
\usepackage{todonotes} 
\usepackage{siunitx} 

\setcopyright{rightsretained}

\acmDOI{}

\acmISBN{}

\acmConference[]{}{}{}
\acmYear{2019}
\copyrightyear{2019}

\acmArticle{4}
\acmPrice{15.00}

\begin{document}

\title{Enhanced Performance and Privacy for TLS over TCP Fast Open}

\author{Erik Sy}
\affiliation{%
  \institution{University of Hamburg}
}

\author{Tobias Mueller}
\affiliation{%
  \institution{University of Hamburg}
}

\author{Christian Burkert}
\affiliation{%
  \institution{University of Hamburg}}

\author{Hannes Federrath}
\affiliation{%
  \institution{University of Hamburg}
}
\author{Mathias Fischer}
\affiliation{%
  \institution{University of Hamburg}
}

\renewcommand{\shortauthors}{E. Sy et al.}

  \begin{abstract}
    {Small TCP flows make up the majority of web flows. For them, the TCP three-way handshake induces significant delay overhead. The TCP Fast Open (TFO) protocol can significantly decrease this delay via zero round-trip time (0-RTT) handshakes for all TCP handshakes that follow a full initial handshake to the same host. However, this comes at the cost of privacy limitations and also has some performance limitations. In this paper, we investigate the TFP deployment on popular websites and browsers. We found that a client revisiting a web site for the first time fails to use an abbreviated TFO handshake in 40\% of all cases due to web server load-balancing using multiple IP addresses.
Our analysis further reveals significant privacy problems of the protocol design and implementation. Network-based attackers and online trackers can exploit TFO to track the online activities of users. As a countermeasure, we introduce a novel protocol called TCP Fast Open Privacy (FOP).
TCP FOP prevents tracking by network attackers and impedes third-party tracking, while still allowing 0-RTT handshakes as in TFO\@.
As a proof-of-concept, we have implemented the proposed protocol for the Linux kernel and a TLS library. Our measurements indicate that TCP FOP outperforms TLS over TFO when websites are served from multiple IP addresses.}
\end{abstract}

%
%



\maketitle

\section{Introduction}

TCP is the standard network protocol for transmitting information on the Internet and a TCP connection is usually used to establish a subsequent TLS connection. Nowadays about 80\% of HTTP web requests are encrypted~\cite{HTTP_Archive}
Retrieving a popular web page requires HTTPS connections to on average 20 different hosts~\cite{sy2019enhanced}, which sums up to 20 TCP and 20 TLS handshakes. Especially for short web flows, these handshakes represents a significant overhead.

To decrease the delay of a TCP handshake, the TCP Fast Open Protocol (TFO)~\cite{rfc7413} has been deployed by the most popular operating systems and browsers, even though it is not yet actively used by all of them.
It shortens TCP's three-way handshake by one round-trip time for all connections that follow an initial TFO connection to the same host.
With the initial TFO handshake, the server verifies the client IP address and sends an identifier (Fast Open cookie) to the client as proof of this verification. The client can then use this cookie for all successive TFO connections to the server as long as its IP address has not changed. However, while providing significant speedup, this identifier can be used to link TFO sessions.
Thus, an online tracker can exploit it to collect profiles of users' browsing behavior.
Furthermore, as TFO messages are sent unencrypted, users can be tracked via passive network monitoring like dragnet surveillance.
Tracking via the TFO protocol is limited as it requires a matching client and server IP address as well as a matching server port for reusing cached Fast Open cookies.

Despite this limitation, tracking via Fast Open cookies can be more effective than tracking based on IP addresses. For example, TFO tracking enables to differentiate between devices sharing the same public visible IP address, e.g., as a result of Network Adress Translation (NAT). Furthermore, TFO tracking allows the attacker to extend the tracking periods compared to IP addresses tracking in cases where the public IP address are dynamically assigned.
Worse, a TFO tracking period is not terminated by a browser restart, nor is it restricted to the scope of a single application on that host, but only by a restart of the kernel or the host (as this clears the kernel's TCP cache).
This is especially worrisome on mobile devices, which are \textit{always on} and are seldomly restarted.
Moreover, TFO tracking is independent of conventional tracking practices such as HTTP cookies or browser fingerprinting~\cite{eckersley2010unique} that use other protocols than IP and TCP\@.

While the most effective countermeasure is to disable TFO entirely, this prevents the round-trip time savings during connection establishment.
To balance the legitimate needs of online privacy and faster TLS over TCP connections, we additionally propose the TCP Fast Open Privacy (TCP FOP) protocol as a countermeasure to TFO tracking.

In summary, this paper makes the following contributions:

\begin{itemize}
  \item To the best of our knowledge, we are the first to describe tracking via TFO cookies. Passive network attackers and online services can use these cookies to link website visits to the same user. We find that the TFO protocol provides no measures to restrict such a tracking mechanism.

  \item We found that under real-world conditions, the first revisit of a website supporting the TFO protocol fails in 40\% of all cases to perform an abbreviated handshake. The main reason for this is server load balancing, i.e., the same website is concurrently served from multiple different IP addresses. This represents a considerable performance limitation of TFO\@.

\item We investigate the TFO configuration of popular browsers and found that the tracking periods for Chrome, Firefox, and Opera seem to be not restricted at all. We successfully tracked successive connections from these browsers for a period of ten days. 
  Furthermore, tracking is feasible for the tested setups across private browsing modes, browser restarts, and even across different browsers running on the same host. Online trackers can utilize TFO to track users as a third-party across multiple websites, and even across different websites, as long as they are served from the same IP address.

\item We propose TCP FOP as a cross-layer solution to overcome the described privacy limitations of TLS over TFO\@. Furthermore, TCP FOP allows abbreviated handshakes for website revisits independently of the server's IP address.
This significantly improves the performance of website revisits compared to TLS over the TFO protocol.
For that, our solution uses an encrypted TLS channel to send Fast Open cookies from the server to the client.
We implemented TCP FOP into the Linux kernel and in a TLS library to demonstrate its real-world applicability. 
The evaluation of our prototype indicates no additional delay compared to TFO/TLS connection establishments.

\end{itemize}

Note that we responsibly disclosed our privacy concerns regarding the TFO protocol to the vendors of popular browsers. As a result of this disclosure, Mozilla deprecated the TFO protocol on all branches of Firefox for all platforms~\cite{Bug_report}.
Furthermore, Microsoft removed support for TFO from the private browsing mode of the Microsoft Edge browser~\cite{edge}.

The remainder of this paper is structured as follows: Section~\ref{sec:TCP-Fast-Open} describes the connection establishment of the TCP Fast Open protocol, its deployment within the Alexa Top Sites, and evaluates its real-world performance limitations. 
Section~\ref{sec:Tracking-TCP-Fast-Open} reviews tracking via TFO cookies, privacy threats arising from host-based as well as network-based attackers, and investigates the feasibility of the presented tracking mechanism for popular browsers. 
Section~\ref{sec:TCP-Fast-Open-Privacy} summarizes TCP FOP as well as its implementation and presents evaluation results. Related work is reviewed in Section~\ref{sec:Related}. Section~\ref{sec:Conclusion} concludes the paper.

\section{TCP Fast Open}\label{sec:TCP-Fast-Open}

In this section, we briefly describe the protocol handshake of TCP Fast Open (TFO).
Subsequently, we investigate the deployment of the TFO protocol for the Alexa Top Million Sites.
We also analyze the performance impact of real-world load-balancing on the rate of zero round-trip time (RTT) connection establishments.
\vspace{-1em}
\subsection{Background on TFO's Connection Establishment}
\begin{figure*}[htpb]
\centering
\begin{minipage}{.3\linewidth}
    \includegraphics[width=\linewidth]{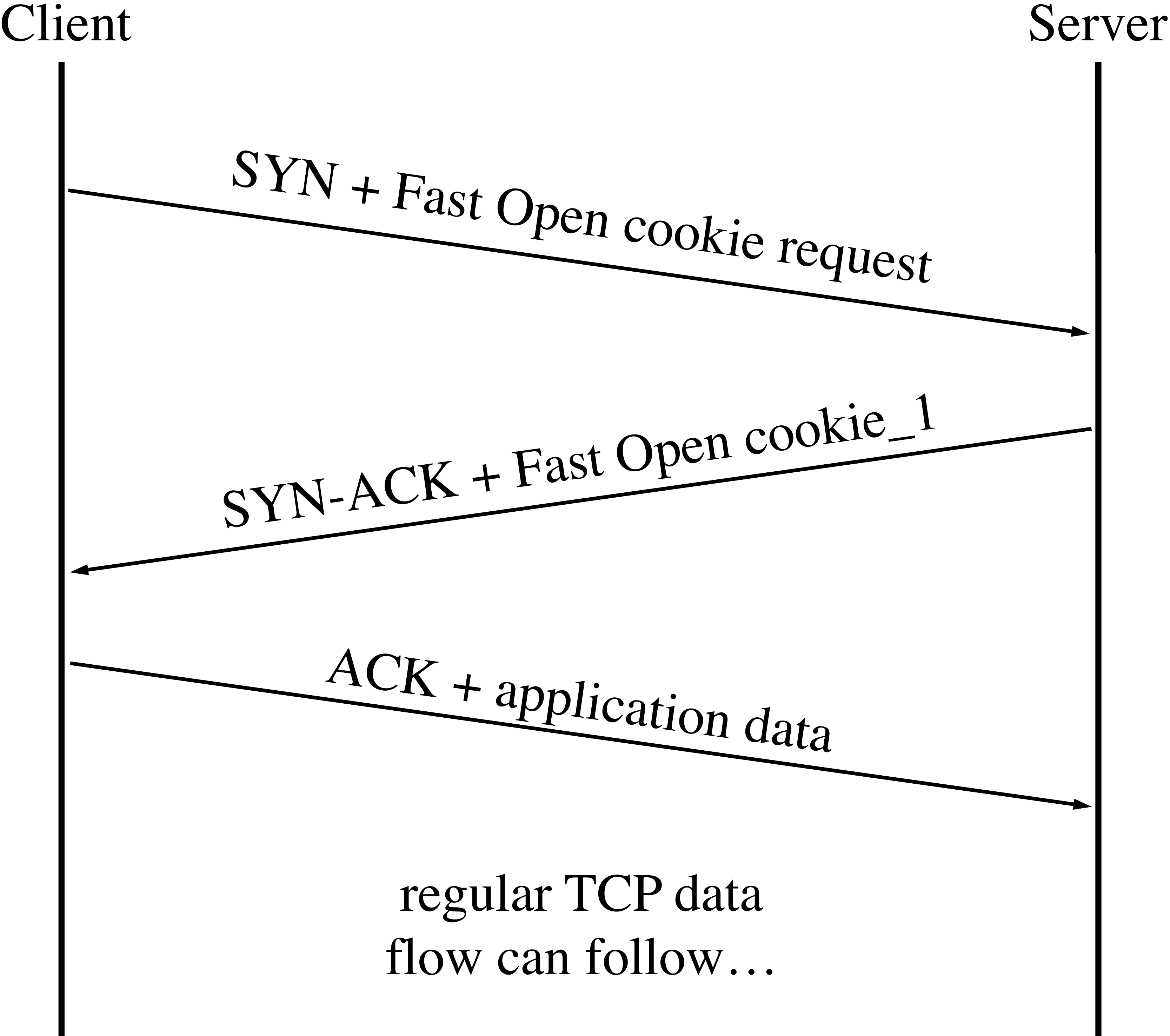}
    \caption*{a) Initial Handshake}
\end{minipage}
\hfill
\begin{minipage}{.3\linewidth}
    \includegraphics[width=\linewidth]{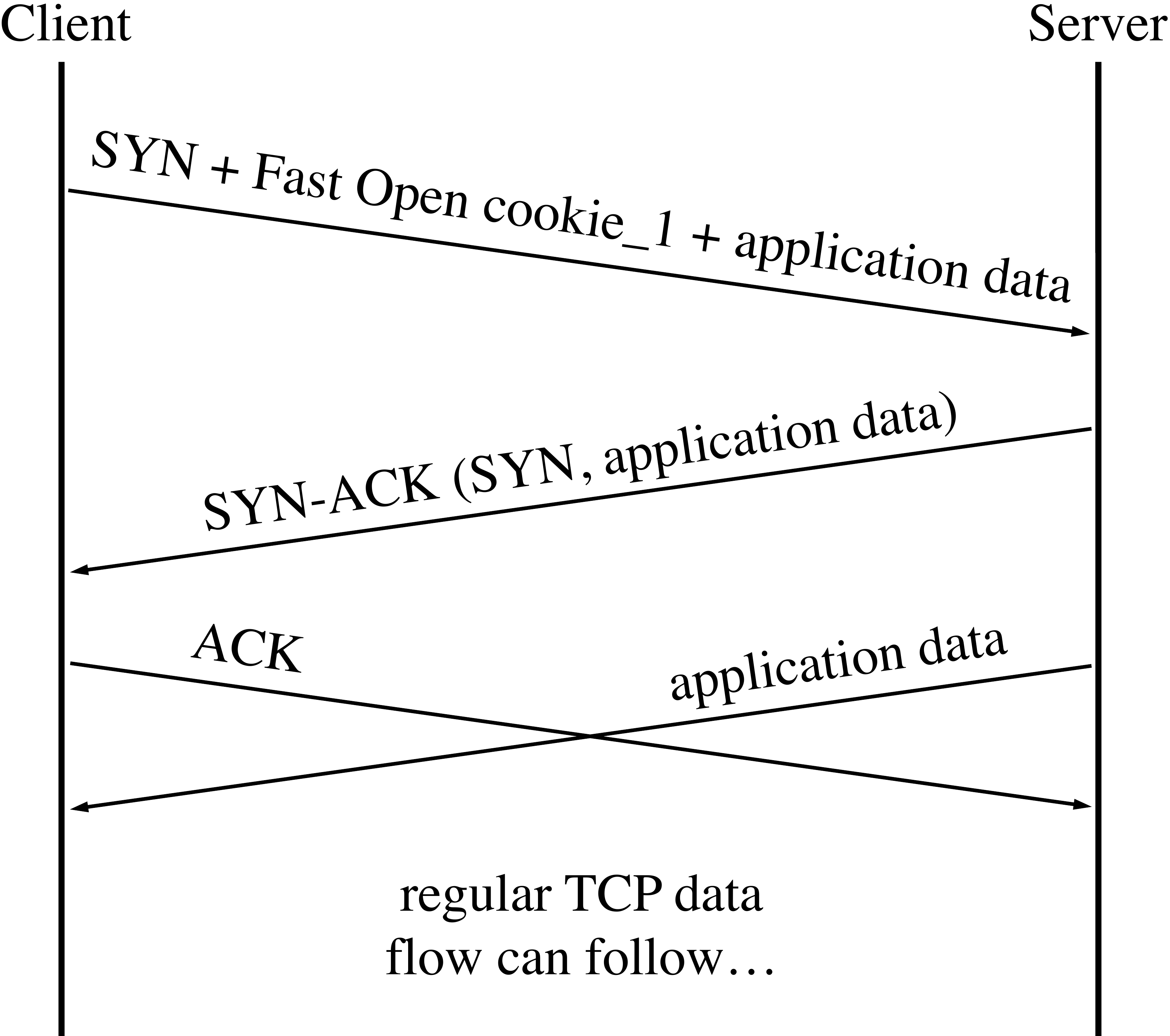}
    \caption*{b) 0-RTT Handshake}
\end{minipage}
\hfill
\begin{minipage}{.3\linewidth}
    \includegraphics[width=\linewidth]{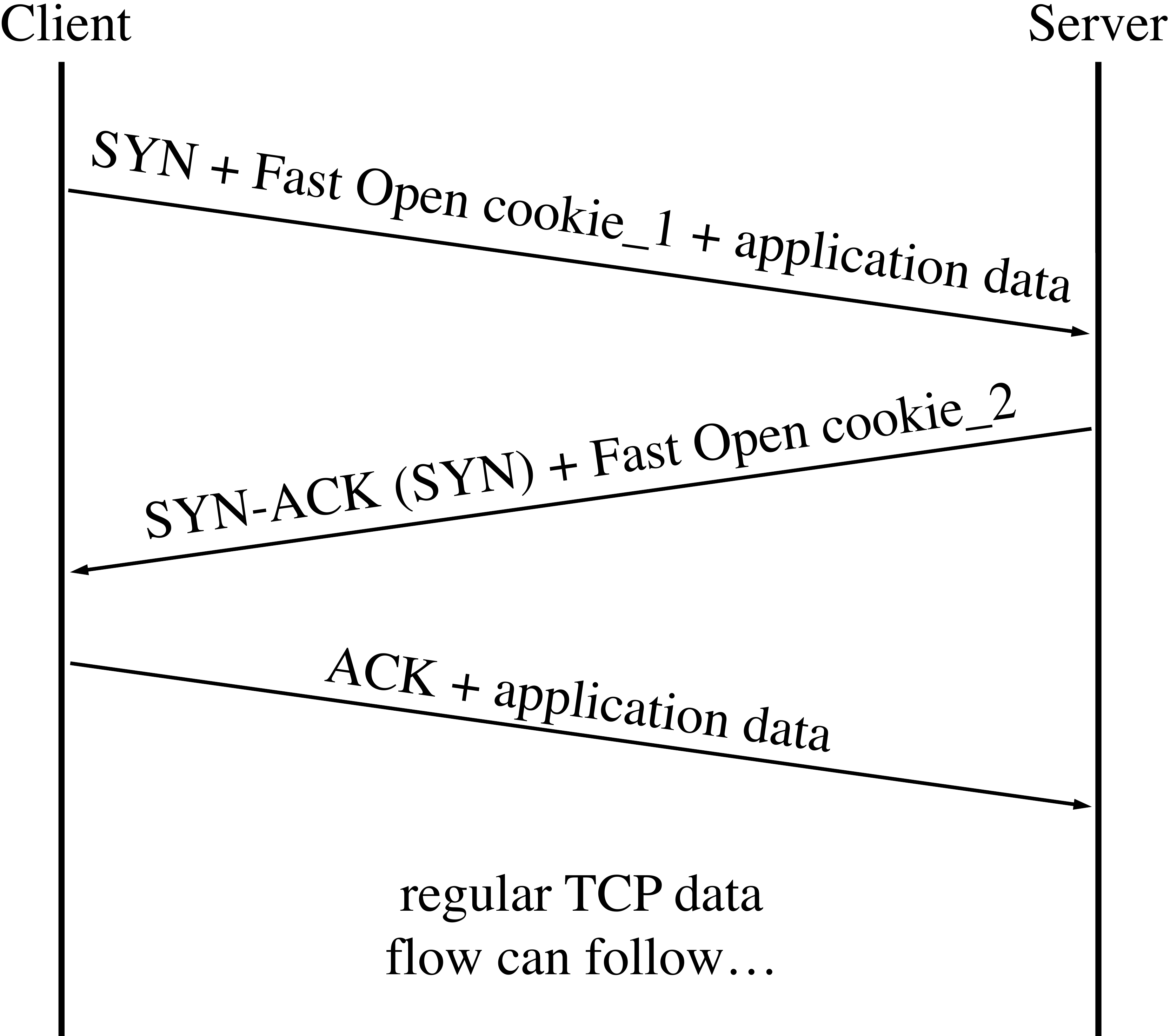}
    \caption*{c) Rejected 0-RTT Handshake}
\end{minipage}
  \caption{Handshakes in TCP Fast Open protocol.}
  \label{fig:TFO_overview}
  \vskip -12pt
\end{figure*}

TFO is defined in RFC 7413~\cite{rfc7413} as an experimental TCP mechanism. It allows saving up to one round-trip time compared to the standard TCP handshake~\cite{rfc793}. 
For that, during a successful TFO handshake the client obtains a cookie from the server, which it can use in subsequent connections. 
The TFO cookie is encrypted and authenticated by the server and opaque to the client.
RFC 7413 does not specify a general construction scheme for these cookies. 
However, it has to contain information about the client's publicly visible IP address.
Thus, if the client presents a cookie which matches its publicly visible IP address, the server accepts this as proof that the client can receive messages at the claimed IP address.
The server then does not need to validate the client's source address via additional message exchanges anymore.
This allows the client to establish a connection without waiting for the server's response.
Thus, application data can be sent immediately along with the first client message. Figure~\ref{fig:TFO_overview} shows a schematic of the TFO handshakes.

\textbf{Initial handshake:} At the beginning, the client has no information about the server.
Similar to a TCP three-way handshake~\cite{rfc793}, the client initiates a connection by sending a SYN to the server as shown in Figure~\ref{fig:TFO_overview}a. 
This SYN includes a TCP option that requests a TFO cookie from the server.
The server confirms the connection request with a message containing a SYN-ACK and a TFO cookie. 
The client then caches the TFO cookie for the establishment of subsequent connections.
To complete the three-way handshake, the client sends an ACK\@.
The now established connection is a standard TCP connection.

\textbf{0-RTT handshake:} For subsequent connections to the same server, the client utilizes the previously retrieved TFO cookie.
For that, it sends the cookie as part of the SYN message to the server as shown in Figure~\ref{fig:TFO_overview}b and c.
Additionally, the client can include application data as payload within the SYN message.
Upon receiving the client's connection request, the server validates the included TFO cookie.

 A cookie is valid for a connection request, if the claimed IP address of the client matches the one associated with the cookie.
For valid cookies, the server accepts the connection request with the attached application data.
As a response, the server sends a SYN-ACK, which acknowledges the client's SYN message and the length of the received application data (see Figure~\ref{fig:TFO_overview}b).
This SYN-ACK message can contain application data as a payload.
In total, this abbreviated connection establishment saves one round-trip time of delay compared to TCP's three-way handshake.

In the case of an invalid cookie, the server drops the application data of the client as shown in Figure~\ref{fig:TFO_overview}c. 
For that, the server sends a SYN-ACK which only acknowledges the client's SYN but not the application data.
Moreover, the server generates a new TFO cookie for the client and attaches this as a payload to the SYN-ACK\@.
Thereafter, the client replaces the cached \textit{cookie\_1} with the fresh \textit{cookie\_2}, which can be used in subsequent 0-RTT handshakes (see Figure~\ref{fig:TFO_overview}c).
Note, that a rejected 0-RTT handshake only causes the same delay as a standard TCP three-way handshake.
\vspace{-1em}
\subsection{Evaluation}

In this section, we first investigate the deployment of TFO within the Alexa Top Million Sites.
This allows us to determine an upper limit of websites which possibly deploy TFO to track their visitors.
Based upon a sample size of approx.\ \num{30000}~hostnames within the Alexa Top Million Sites, we then investigate to which extent changing server IP addresses affects the performance gains achievable by TFO\@.
\vspace{-1em}
\subsubsection{Deployment of TFO}\label{sec:deploy}

Major operating systems such as Windows, macOS, Linux, FreeBSD, Android, and iOS support the TFO protocol, which is a precondition for its widespread adoption.
However, these implementations do not set TFO as a default for all TCP connections and thus it still requires modifications to the client- and server-side applications to be used.
The RFC describing the TFO protocol was published in 2014~\cite{rfc7413}.
We thus assume that our measurement of the TFO deployment investigates an early-stage in the wide-spread adoption of this protocol that we expect in the near future.

To approximate the deployment of TCP Fast Open on the Internet, we investigate the support for TFO within the Alexa Top Million Sites~\cite{Alexa}.
For this purpose, we sent a SYN packet containing a TFO cookie request to the first IP address mentioned in the DNS record of the corresponding hostname.
In this measurement, we did not treat hostnames hosted by Content Delivery Networks (CDNs) differently.
Our measurements succeeded to receive a response from a server for 97.1\% of the investigated hostnames.
The remaining 2.9\% of our measurements resulted in errors.
In detail, we observed errors in the name resolution for \num{19498} domain names.
Furthermore, we did not receive a response to our TFO connection request from \num{9254} hosts.
If the respective host responded with a SYN-ACK including a TFO cookie, we consider this site to support the TFO protocol and the contrary otherwise.
We limited our scans to port 443 on the targeted web server as this is the standard port for HTTPS web services~\cite{rfc1700}.
We conducted this measurement from an IPv4 address on the 10th of August 2018 using a dedicated Python script.

 \begin{table}[htbp]
   \caption{Websites with TFO-support in Alexa Top lists}
  \label{tab:alexa_lists}
 \centering
 \begin{tabular}{lr}
\toprule 
\multicolumn{1}{c}{Alexa Top lists} & \multicolumn{1}{c}{Share of hostnames with TFO-support} \\
 \midrule
Alexa Top 10&60.0\%\\
 Alexa Top 100 &28.0\%\\
 Alexa Top 1K&12.4\%\\
 Alexa Top 10K &5.9\%\\
 Alexa Top 100K&3.4\%\\
 Alexa Top 1M &3.2\%\\
 \bottomrule 
 \end{tabular}
 \end{table}
Table~\ref{tab:alexa_lists} shows the number of hostnames supporting the TCP Fast Open protocol within different Alexa Top lists.
We find that 60\% of the ten most popular hostnames support the protocol.
However, this fraction decreases with the size of the Alexa Top list.
While 28\% among the Top Hundred hostnames still enable TCP Fast Open handshakes, this share decreases to 3.2\% within the Top Million sites.
We assume that higher-ranked websites tend to adopt new protocols such as TCP Fast Open earlier than other websites.
\vspace{-1em}
\subsubsection{Performance Limitations of TFO} \label{sec:performance}

\begin{figure}[tbp]
\centering
\includegraphics[width=0.38 \textwidth]{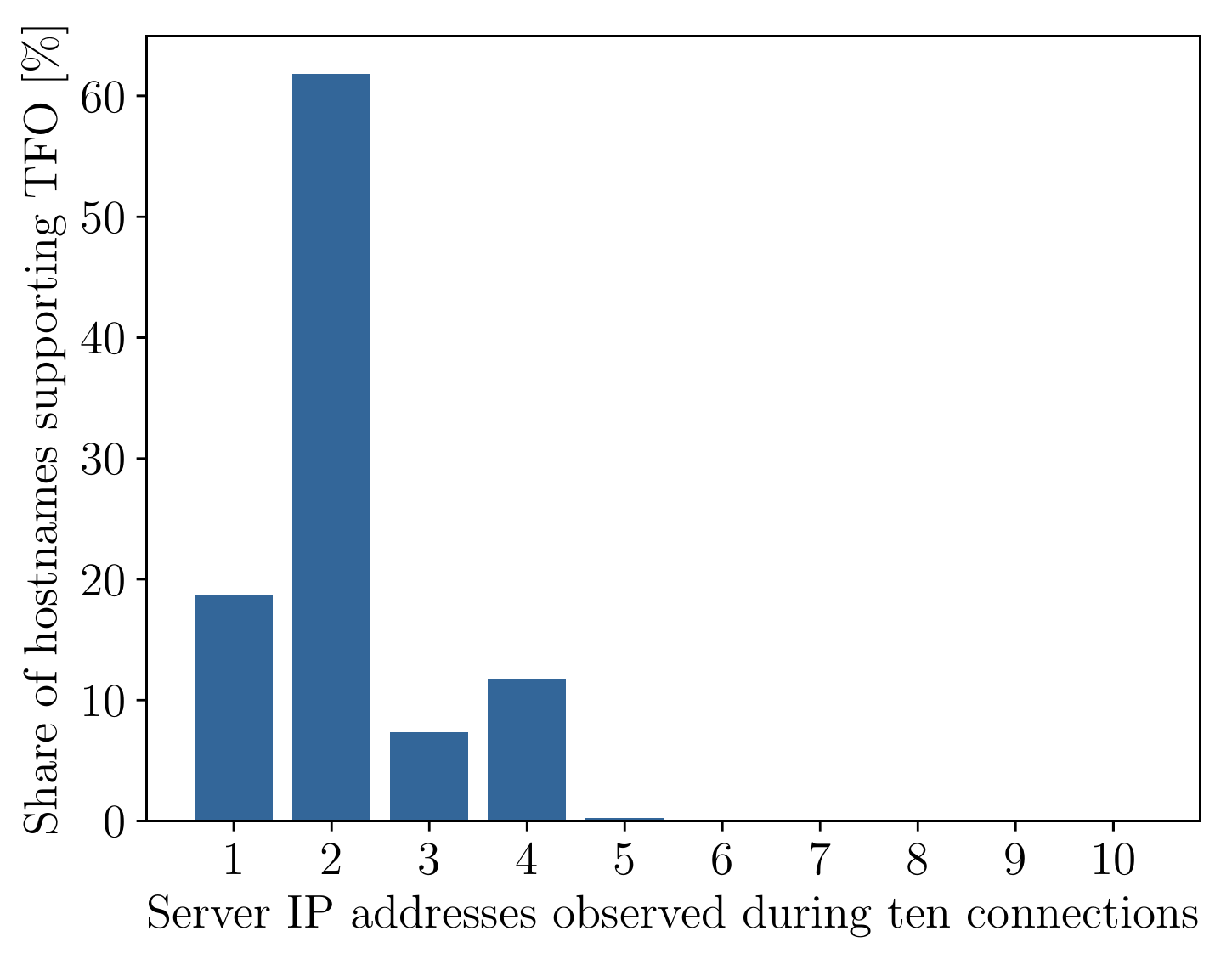}
\caption{Share of hostnames with TFO-support plotted over the number of observed server IP addresses for ten connections.}
\label{fig:ip-changes_count}
\vskip -12pt
\end{figure}
Repeated connections to a hostname are not necessarily served from the same IP address due to server load balancing.
However, the TFO protocol instructs to utilize a cached Fast Open cookie only if the source IP address, destination IP address, and the destination port match those of the TCP connection in which the cookie was issued.
As a result, any time the hostname is resolved to a different IP address, the client experiences a cache miss even if a Fast Open cookie from that hostname is stored in the TCP cache.
To assess the performance impact of this design, we observe the IP addresses of the responding servers, while connecting to a hostname several times. 
We conducted this measurement on August 24th, 2018 using a virtual machine in the data center of our university.
We used a dedicated Python script to send a TFO connection request using a static IPv4 address.
Subsequently, we observed the server's response.
When the responded SYN-ACK includes a TFO cookie, we consider this host to support TFO and the contrary otherwise.
Note, that we did not attempt handshakes using a previously retrieved cookie.
In total, we connected to \num{32099}~hostnames that we identified in our previous measurement (see Section~\ref{sec:deploy}) as sites supporting the TFO protocol. 
We connected to every hostname ten times with intervals of 45~minutes between each successive connection and observed the respective server IP addresses.
This time interval was chosen for reasons of convenience because our test setup required about 40~minutes to initiate connections to the entire set of investigated hostnames.
Note, that the selected time interval is significantly larger than the expiration time of DNS records of popular websites, which mostly expire within 5~minutes~\cite{sy2019resolver}.
During this measurement, we experienced a failure rate of 1.2\% that was mainly caused by unresponsive hosts. 

\begin{figure}[tbp]
\centering
\includegraphics[width=0.38 \textwidth]{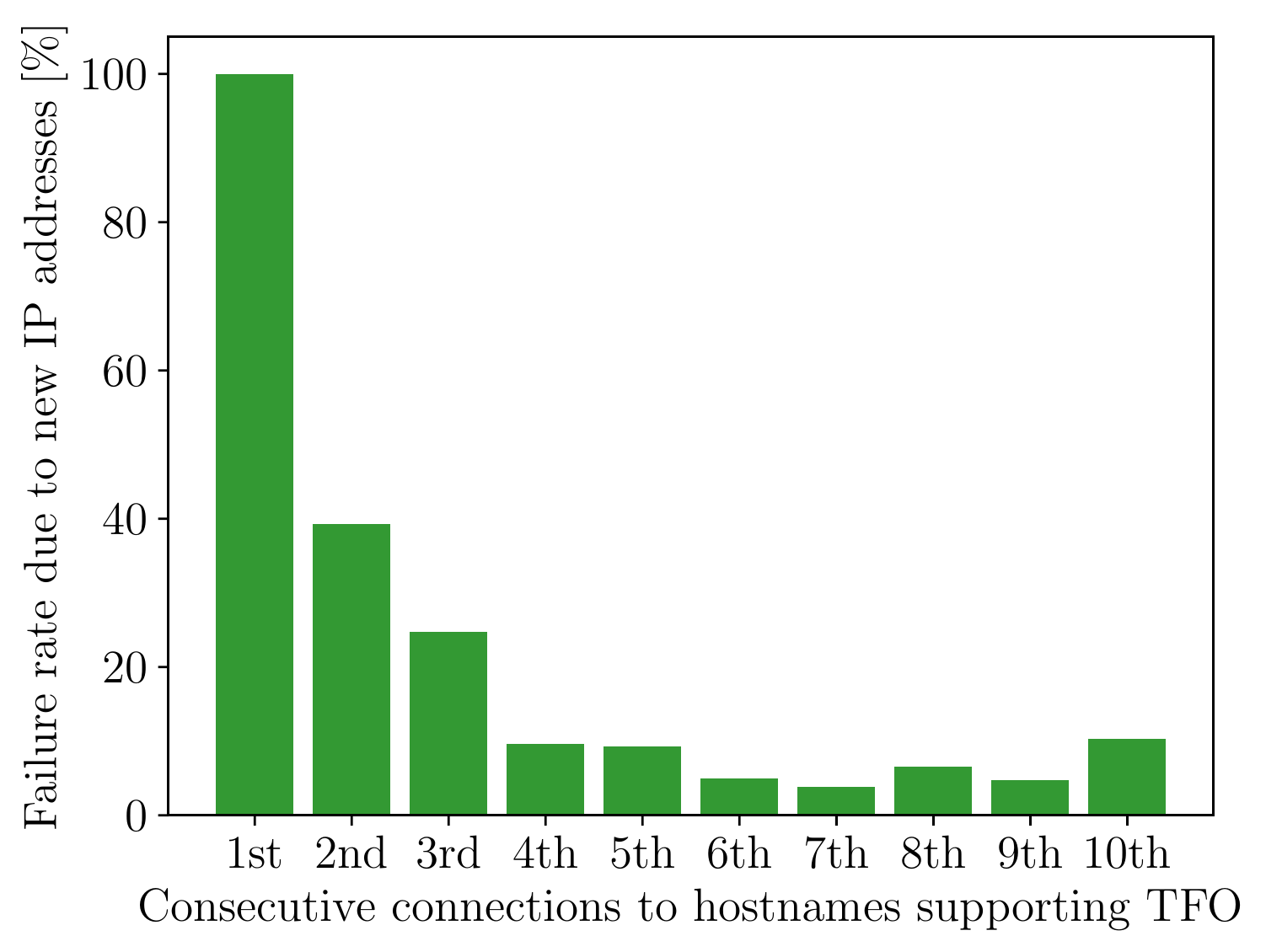}
\caption{Failure rate to conduct abbreviated handshakes due to fresh server IP addresses plotted over successive connections to hostnames with TFO-support.}
\label{fig:failure_rate}
\vskip -12pt
\end{figure}

About 94\% of the investigated hostnames indicated, support for the TFO protocol at each of these connections.
The remaining 4.8\% can be attributed to hostnames that are served from multiple servers and not all of them support the TFO protocol.
This includes servers not supporting TFO on a different IP address, but also on the same address in case anycast is used to reach multiple physical servers behind the same address.
Figure~\ref{fig:ip-changes_count} shows the number of observed IP addresses per site for the \num{30218}~hostnames always supporting TFO\@. 
Our results indicate, that at least 81\% of the tested hostnames are served from several IP addresses.
We also found that the second connection to a hostname introduces \num{11876} new IP addresses compared to the first connection.
As shown in Figure~\ref{fig:failure_rate}, the second TFO connection to a hostname fails in 39.3\% of all attempts due to the fact that the hostname is served from a different IP address than the first.
For the third connection to a hostname, we observe an average failure rate of 24.7\%.
On average, the tested hostnames are served each from 2.1 different IP addresses across our measurements.
We conclude that the binding of TFO cookies to a specific server IP address presents a significant real-world performance limitation in the context of web browsing.

\section{Tracking via TCP Fast Open}%
\label{sec:Tracking-TCP-Fast-Open}

In this section, we begin by introducing a basic approach to track users via the TFO protocol.
Based on that, we describe practical tracking scenarios using this approach.
Afterward, we compare TFO-based tracking to IP-based tracking.
Finally, we evaluate the default configuration of popular web browsers regarding TFO to assess the real-world impact of the presented tracking mechanism. 

\vspace{-1em}
\subsection{Basic Tracking Approach}

Essential to TFO-based tracking are the cookies that the TFO protocol uses to authenticate a client upon consecutive connections to a server.
These cookies are generated by the server and permit the attacker to identify clients with up to 16~bytes of entropy~\cite{rfc7413}.
They enable the server to link all connections in which the same unique cookie is used and to attribute them to the same client.
Moreover, a failed authentication as shown in Figure~\ref{fig:TFO_overview}c allows the server to link the fresh \textit{cookie\_2} to the same client that was previously using \textit{cookie\_1}. 
Following TFO's specification, the client should then use the fresh \textit{cookie\_2} for subsequent connections to the same server.
In total, the server can track its clients via Fast Open cookies in an essentially similar fashion as with the widespread HTTPS cookie.

\vspace{-1em}
\paragraph*{Attacker Model}
Our attacker model assumes that
the attacker can read network packets including their IP and TCP headers. 
The attacker is capable of extracting Fast Open cookies from the TCP headers and storing them for future reference in association with the respective tracking profile.
As a limitation, the attacker is not capable to conduct user tracking based on any other protocol except for IP and TCP\@.
The attacker cannot break cryptographic primitives. Hence, they cannot violate the confidentiality or authenticity of the TFO cookie without access to the respective secret key.
The attacker has no direct access to a client and cannot violate the integrity of the software run by the client.
Additionally, a host-based attacker located on the server has access to the secret key used to encrypt and authenticate TFO cookies and can thus issue custom TFO cookies.



\vspace{-1em}
\subsection{Tracking Scenarios}

While tracking by a single web server presents a privacy issue in itself, it is amplified if a tracker can identify a user across several visited websites.
The remainder of this section, describes third-party tracking, tracking across virtual domains, and tracking by a network-based attacker, all of which allow tracking via the TFO protocol across multiple websites.

\vspace{-1em}
\paragraph*{Scenario 1: Third-party Tracking}

Third-party tracking refers to a practice, where a party other than the targeted website can link website visits to the same user.
This is a common practice on the Internet considering that Alexa Top~500 websites include on average 17.7~third-party trackers~\cite{englehardt2016online}.
The presented tracking mechanism allows identifying users across all websites where a corresponding tracker is included as a third-party resource.
However, to distinguish the various first-party sites, i.e.,\ referrers, that a user visited, the tracker requires an additional identifier such as an HTTP referrer or a dedicated URL per first-party.

\vspace{-1em}
\paragraph*{Scenario 2: Tracking Across Virtual Domains}

In virtual hosting, multiple virtual domains are hosted on a single server or pool of servers.
This approach allows sharing resources like the IP address and server hardware across domains.
When domain name~$\mathcal{A}$ and $\mathcal{B}$ share the same IP address, then a TFO connection to both websites will contain the same cookie.
Hence, an operator of a virtual hosting platform such as a Content Delivery Network (CDN) can link visits of the same user across the hosted virtual domains.
In detail, a Fast Open cookie issued during a connection to domain name~$\mathcal{A}$ will be reused by the client when connecting to domain name~$\mathcal{B}$ allowing the operator of these hosts to link these connections  based on the same Fast Open cookie to the same user.

\vspace{-1em}
\paragraph*{Scenario 3: Tracking by a Network-based Attacker}

Since TCP itself provides no confidentiality, a passive, network-based attacker can observe the content of TCP headers as long as no protective measure is taken on lower protocol layers, e.g., IPsec.
As a consequence, the attacker can use TFO cookies for tracking purposes as described above.
This is particularly sensitive if the network-based attacker is located on the public Internet and targets a client that is situated in a local network which uses address translation (NAT). In that case, the attacker would be otherwise unable to distinguishing specific users.
As a result, this undermines the efforts of protocols such as TLS 1.3~\cite{rfc8446} that aim to protect against tracking by network-based attackers.

%
\vspace{-1em}
\subsection{Comparison to Tracking via IP Addresses}

In this section, we describe how the presented mechanism extends the capabilities of tracking via IP addresses despite TFO's requirement for matching client and server IP addresses to reuse cached Fast Open cookies.
To illustrate this comparison, we introduce a scenario of devices sharing the same IP address and another scenario where a device is placed behind a Network Address Translator (NAT), whose publicly visible IP address changes dynamically.
\vspace{-1em}
\subsubsection{Distinguishing Devices Sharing an IP Address}
The Internet is dominated by the use of IP version~4 and the available address space is already exhausted~\cite{IPv4_exhaust}.
This makes the sharing of IPv4 addresses necessary.
Furthermore, the transition to IP version~6 is difficult and it is only supported by about a quarter of the most popular \num{10000} websites and the global Internet users, respectively~\cite{sy2019resolver, IPv6_user}.
Thus, the sharing of IP addresses is a common scenario from small home networks up to large carrier-grade NATs spanning several thousand devices.
Tracking based on IP addresses becomes infeasible when a large number of devices share the same IP address as they form a large anonymity set.
Tracking via Fast Open cookies does not have the same limitation.
The tracker can provide each device with a unique Fast Open cookie.
Thus, a client presenting this cookie during a future connection request can be clearly linked to the prior connection that was used to issue the cookie.
\vspace{-1em}
\subsubsection{Prolonged Identification behind NAT}
\begin{figure}
\centering
\includegraphics[width=0.4 \textwidth]{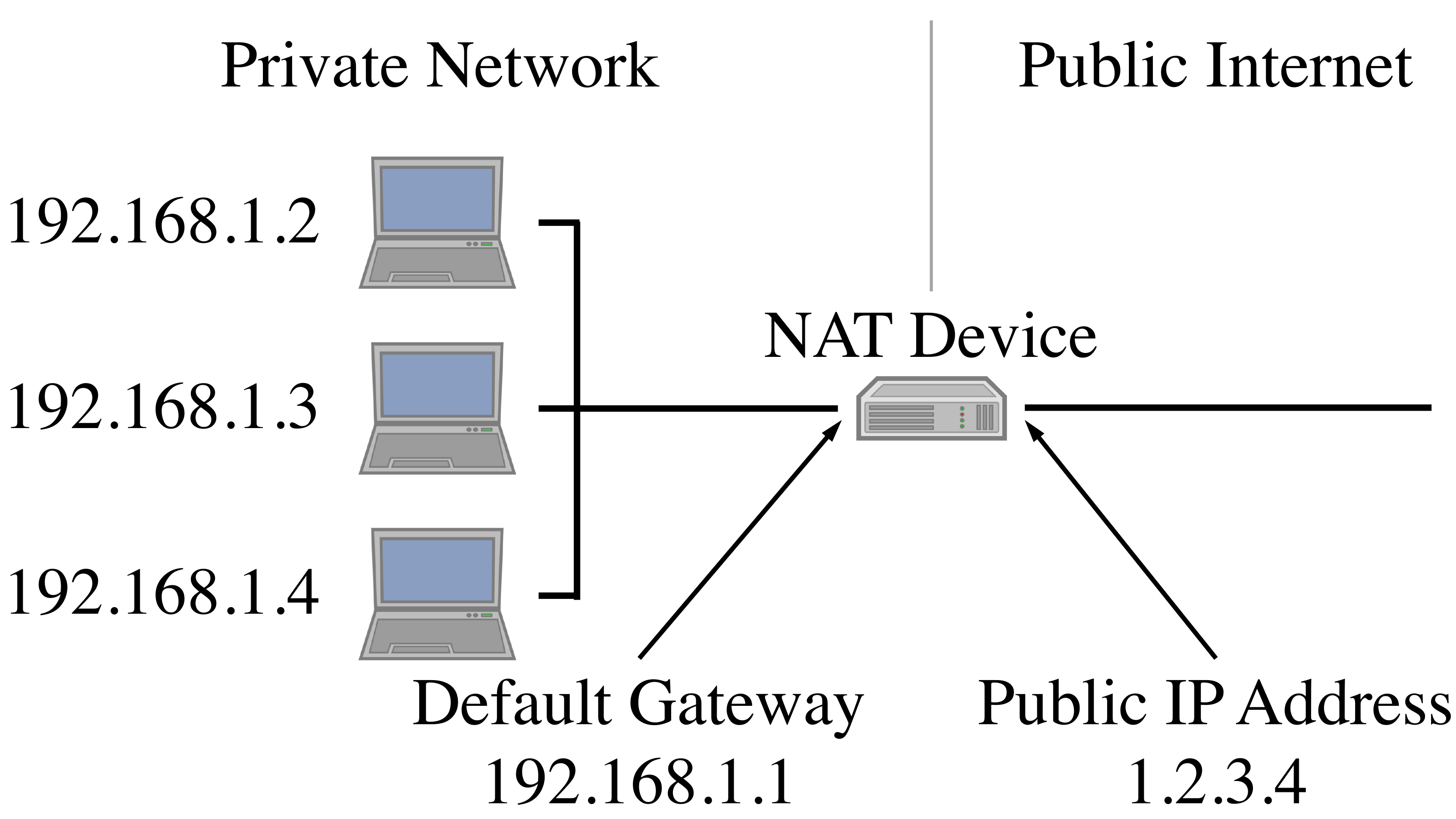}
\caption{NAT device used for a private network.}
\label{fig:NAT}
\vskip -12pt
\end{figure}

On the Internet, a significant share of users uses a dynamically assigned IP address~\cite{Xie:2007}.
Figure~\ref{fig:NAT} shows a common network topology, where a user's device is located in a private network behind a NAT gateway that always uses the same local IP address.
Furthermore, we assume that the public IP address of the NAT gateway changes dynamically.
In this setup, user tracking via IP addresses stops when the NAT gateway receives a new public IP address because a tracker on the Internet cannot link the old to the new IP address.
In the case of TFO, the client's static, local IP address always fulfills the requirement of a matching client IP address during the reuse of cached cookies.
Thus, the client device reuses cached TFO cookies independently of its publicly visible IP address assigned to the NAT\@.
In total, in the described setup, tracking via TFO cookies allows extending the feasible tracking periods compared to when using only the client's publicly visible IP address.

Note, that the tracking periods achievable via TFO are limited by the uptime of an OS, as a restart clears the corresponding TCP cache.
However, especially mobile devices such as smartphones can achieve an uptime of several days or weeks under real-world conditions and thus allow for tracking periods of a similar duration.
Another limiting factor is the TCP cache size, which under certain circumstances can lead to the eviction of older entries~\cite{tcp_metrics}.

\vspace{-1em}
\subsection{Evaluation}\label{sec:browser_eval}

To explore the real-world feasibility of user tracking via the TFO protocol, we investigated popular web browsers on different operating systems.
In total, we conducted eight browser experiments, whose methodology and results we present in the remainder of this section.
\vspace{-1em}
\subsubsection{Status of TFO on the test systems}

The privacy problems of TFO are relevant to all applications that use this protocol. 
However, within the scope of this work, we focus our investigation on popular web browsers because of their important role to protect users' web browsing behavior against online tracking without user consent.
In our sample of popular web browsers, we included the Top~3 mobile and the Top~6 desktop browsers~\cite{browser_desktop_market}.
We tested those browsers on up-to-date versions of Android, iOS, Linux, macOS, and Windows~10 and investigated their support for the TFO protocol by analyzing the network traffic between browser and server.
We find, that the deployment of the TFO protocol in popular browsers is still at an early stage.
Thus, only Microsoft Edge on Windows~10 supports TFO by default. 
Firefox, Chrome, and Opera support the TFO protocol as an experimental feature on several operating systems as shown in Table~\ref{tab:browser}.
Note, that TFO is activated by default within Firefox Nightly and Firefox Beta under macOS and Windows~10, which indicates preparations to further deploy TFO across the Firefox platforms. 
Our tests for iOS~11 and Android Kernel 4.10 did not reveal any popular browser which supports TFO\@.
Note, that the TFO implementation of Microsoft Edge did not work reliably within the IPv4 network stack. To overcome this issue, we tested this browser with an IPv6 network stack, while all other test systems deployed IPv4.
 
  \begin{table*}[htbp]
   \caption{TCP Fast Open default configuration of popular browsers. Note, that the experiments with the Edge browser only worked reliably with IPv6, while all other setups were tested with IPv4. Due to the feature of temporary IPv6 addresses, the default Windows 10 behavior issues a fresh temporary IPv6 address every 24 hours, which is then used by the Edge browser.}
  \label{tab:browser}
 \centering
 \begin{tabularx}{\linewidth}{ll*{7}{>{\centering\arraybackslash}X}}
 \toprule
 \multicolumn{1}{c}{} & \multicolumn{1}{c}{} & \multicolumn{1}{c}{} &  \multicolumn{6}{c}{Tracking across}  \\
 \cmidrule(){4-9}
Browser/Test system & Status &  Tracking periods & Third-parties & Virtual hosts & IP addr.\ changes & Private browsing modes & User applications & Browser restarts \\
 \midrule
Chrome v68/Ubuntu 18.04&support&unrestricted&viable&viable&blocked&viable&viable&viable\\
Firefox v61/Ubuntu 18.04&support&unrestricted&viable&viable&blocked&viable&viable&viable\\ 
Firefox v61/macOS 10.13&support$^*$&unrestricted&viable&viable&blocked&viable&viable&viable\\ 
Firefox v61/Windows 10&support$^*$&unrestricted&viable&viable&blocked&viable&viable&viable\\ 
Edge v42/Windows 10&default&24 hours&viable&viable&blocked&viable&viable&viable\\ 
Opera v54/Ubuntu 18.04&support&unrestricted&viable&viable&blocked&viable&viable&viable\\ 
\bottomrule
 \end{tabularx} 
 \vskip 6pt
  \footnotesize{$^*$Activated by default within Firefox Nightly and Firefox Beta.}\\
 \end{table*}
\vspace{-1em}
\subsubsection{Feasible tracking periods}
This measurement gives a lower boundary of feasible tracking periods via the presented approach.
For that, we visited a website that supports TFO\@.
In between different visits, we closed the browser tab that was in use and left the browser idle in the background of the operating system. 
After one hour, we attempted to revisit the same website served from the same IP address.
By a manual analysis of the network traffic between the browser and the server we observed whether the browser attempted to use the cached Fast Open cookie from the first website visit to establish the fresh connection. 
If the browser makes use of the cookie, we can then assume that it is feasible to track users with the deployed test setup and for the duration of our test.
Next, we repeated this measurement with a fresh TCP cache and increasing the delays between consecutive website visits up to ten days.
We found that none of the IPv4-based test setups indicated a restriction of the feasible tracking period.
Thus, we could track all Chrome, Firefox, and Opera setups for the entire test period of ten days as shown in Table~\ref{tab:browser}.
Furthermore, this indicates that the tested operating systems do not limit the feasible TFO tracking periods via the TFO protocol.
For the Microsoft Edge browser, we were required to conduct this experiment on an IPv6 network stack, which diverts from our other browser test setups.
We observed for the Windows 10 default configuration that the Edge browser utilizes temporary IPv6 addresses~\cite{rfc4941} within the available address block.
As a privacy feature, the lifetime of these temporary addresses is limited to 24~hours by Windows~10.
Thus, this test setup changes its global IPv6 address after 24~hours, even when the assigned IPv6 address block remains the same.
However, the TFO protocol only uses cached Fast Open cookies if the source IP address of the test system is the same as in the TFO connection from which the cookie was retrieved.
As a result, the observed tracking periods terminate with the change of the temporary IPv6 address.
\vspace{-1em}
\subsubsection{Tracking across third-parties}

\begin{figure}
\centering
\includegraphics[width=0.4 \textwidth]{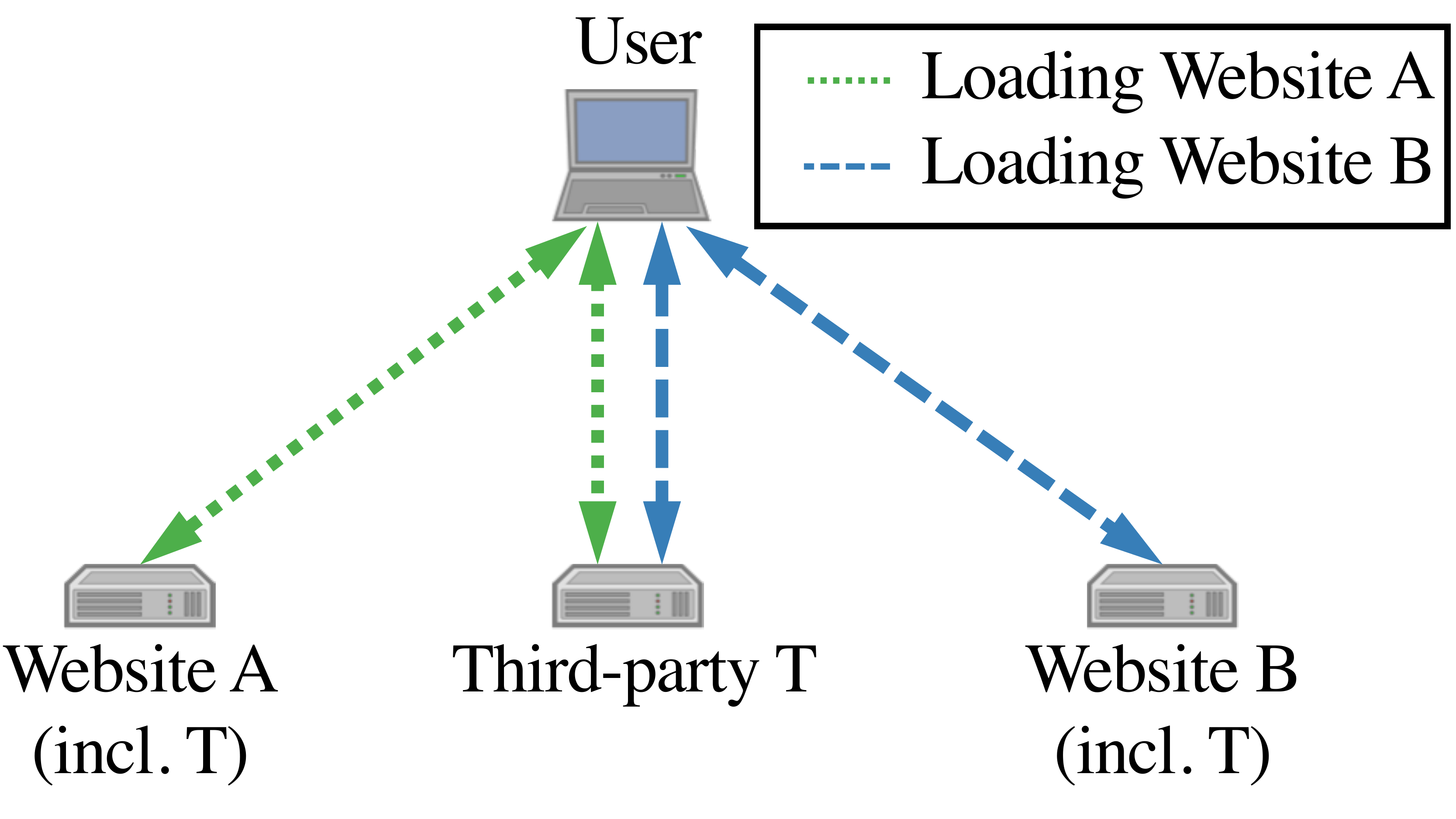}
\caption{Testbed to measure browser behavior regarding third-party tracking.}
\label{fig:third-party}
\vskip -12pt
\end{figure}

This measurement investigates the feasibility of third-party tracking via the TFO protocol.
Figure~\ref{fig:third-party} shows a schematic of the deployed test setup for this experiment. 
To conduct this measurement, we require two websites $\mathcal{A}$ and $\mathcal{B}$ that include the same third-party $\mathcal{T}$.
We visited website $\mathcal{A}$ and validated that the browser received a Fast Open cookie from the third-party $\mathcal{T}$ by manually analyzing the network traffic.
After closing the browser tab in-use, we waited for 30 minutes for open TFO connections to time out~\cite{tcp_man}.
We then visited website $\mathcal{B}$ and investigated the network traffic between the browser and the third-party  $\mathcal{T}$.
If the browser attempted to use the cached Fast Open cookie for the connection establishment with $\mathcal{T}$, we concluded that third-party tracking via TCP Fast Open is feasible with this browser. 
As shown in Table~\ref{tab:browser}, none of the tested browsers applied mechanisms to prevent third-party tracking via Fast Open cookies.
Thus, third-parties present on several websites can track the same user's visits across all those sites.
\vspace{-1em}
\subsubsection{Tracking across virtual hosts}

This experiment is used to investigate the feasibility of tracking across virtual hosts.
It requires two websites whose DNS entries direct to the same IP address.
We connected to one of these websites and afterward closed the browser tab and waited for 30 minutes to ensure that TCP connections to that website timed out.
We then connected to the second website and monitored the respective network traffic of this connection.
If the second connection uses the Fast Open cookie, which was retrieved during the connection to the first website, we conclude that tracking across virtual hosts is feasible with the tested browser. 
Our evaluation indicates that the investigated browsers do not prevent tracking across virtual hosts (see Table~\ref{tab:browser}).
Therefore, when multiple websites are served from the same IP address, the service operator can identify users across those hosted websites.
\vspace{-1em}
\subsubsection{Tracking across IP address changes}

This test investigates the browser behavior regarding TFO when the operating system gets a new IP address assigned.
To assess this behavior, we visited a website and closed the browser tab afterward.
While the browser was running idle in the background of the operating system, we assigned a new IP address to the device.
Then, we revisited the website with the tested browser instance and monitored the network traffic of the connection.
If the browser reuses the cached Fast Open cookie of the website, we can conclude that tracking across IP address changes is feasible.
We observed, that user tracking is not feasible across IP address changes of the client as shown in Table~\ref{tab:browser}.
However, considering a common consumer setup, where devices reside in a private network that is connected to the Internet through a NAT gateway, such devices typically keep their local IP addresses unchanged indefinitely, since DHCP servers deterministically reassign the same local IP address based on unchanging features like the client's MAC address.
In such a setup, the client's unchanging local IP address is independent of the public IP address, which is assigned to the NAT gateway.
Thus, even after a change of the public IP address, the client will try to connect using the previously cached Fast Open cookies, which are bound to the unchanged local sender address.
As a consequence, this allows a tracking server to learn a client's new publicly visible IP address and continue its tracking activities across the IP address change.
\vspace{-1em}
\subsubsection{Tracking across private browsing modes}
This experiment explores whether user tracking via TFO is feasible across browsers' default and private browsing mode.
To assess this browser behavior, we visited a website in the default mode of a browser.
Then, the respective browser tab was closed and the private browsing mode activated.
While monitoring the network traffic, we then revisited the website.
If the connection establishment in the private browsing mode used the previously retrieved Fast Open cookie, we could conclude that tracking across the browsing modes of the tested browser is feasible.
As indicated in Table~\ref{tab:browser}, all setups allow a remote online tracker to identify their user across changes of their browsing mode.
This observed behavior presents a breach in the respective privacy modes, which aim to discard cookies at the end of each private session~\cite{private_mode}.
\vspace{-1em}
\subsubsection{Tracking across browser restarts}

This measurement tests whether tracking across browser restarts is feasible.
To investigate this browser behavior, we first visit a website and retrieve a fresh Fast Open cookie.
Then, we restart the browser and revisit the same website while we monitor the respective network traffic.
If the browser reuses the cookie of the previous browser instance, then we conclude that tracking across browser restarts is feasible with the deployed setup.
We find, that none of the tested browsers prevents tracking via Fast Open cookies across a browser restart.
\vspace{-1em}
\subsubsection{Tracking across user applications}

This experiment explores tracking across different user applications on the same device. 
To conduct the experiment, we retrieve a website and leave the respective browser idle in the background of the OS\@. 
Afterward, we use another application with support for the TCP Fast Open protocol such as another browser or curl to retrieve the same website. 
By monitoring the network traffic between the website and the operating system, we find out whether the second application reuses the TCP Fast Open cookie of the tested browser. 
If so, then tracking across user applications is feasible.
Our results indicate, that user tracking across applications on the same client operating systems is viable for all tested setups (see Table~\ref{tab:browser}).

\begin{figure*}[htpb]
\centering
\begin{minipage}{.35\linewidth}
    \includegraphics[width=\linewidth]{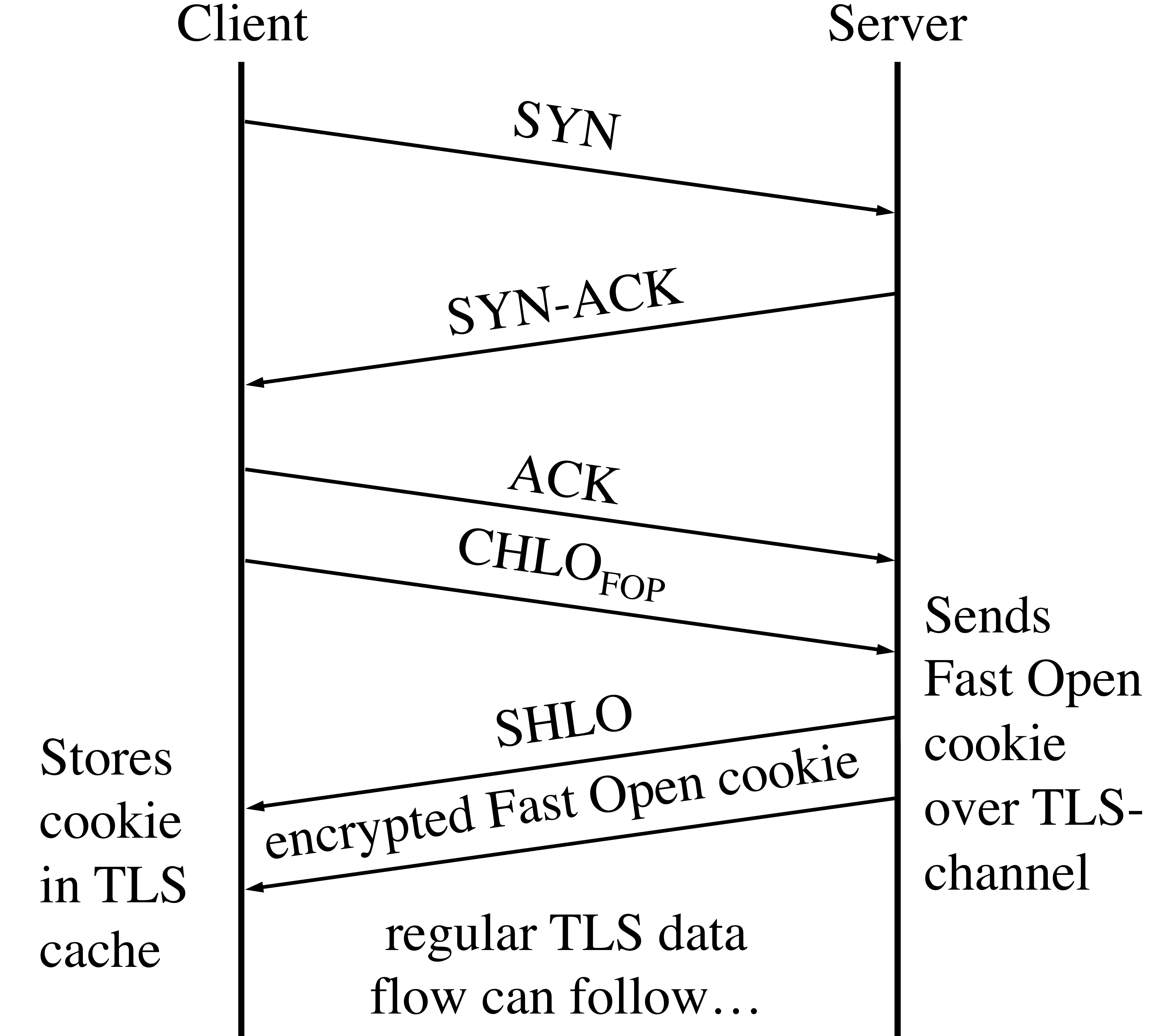}
    \caption*{a) Initial Handshake}
\end{minipage}
\hspace{1.5cm}
\begin{minipage}{.35\linewidth}
    \includegraphics[width=\linewidth]{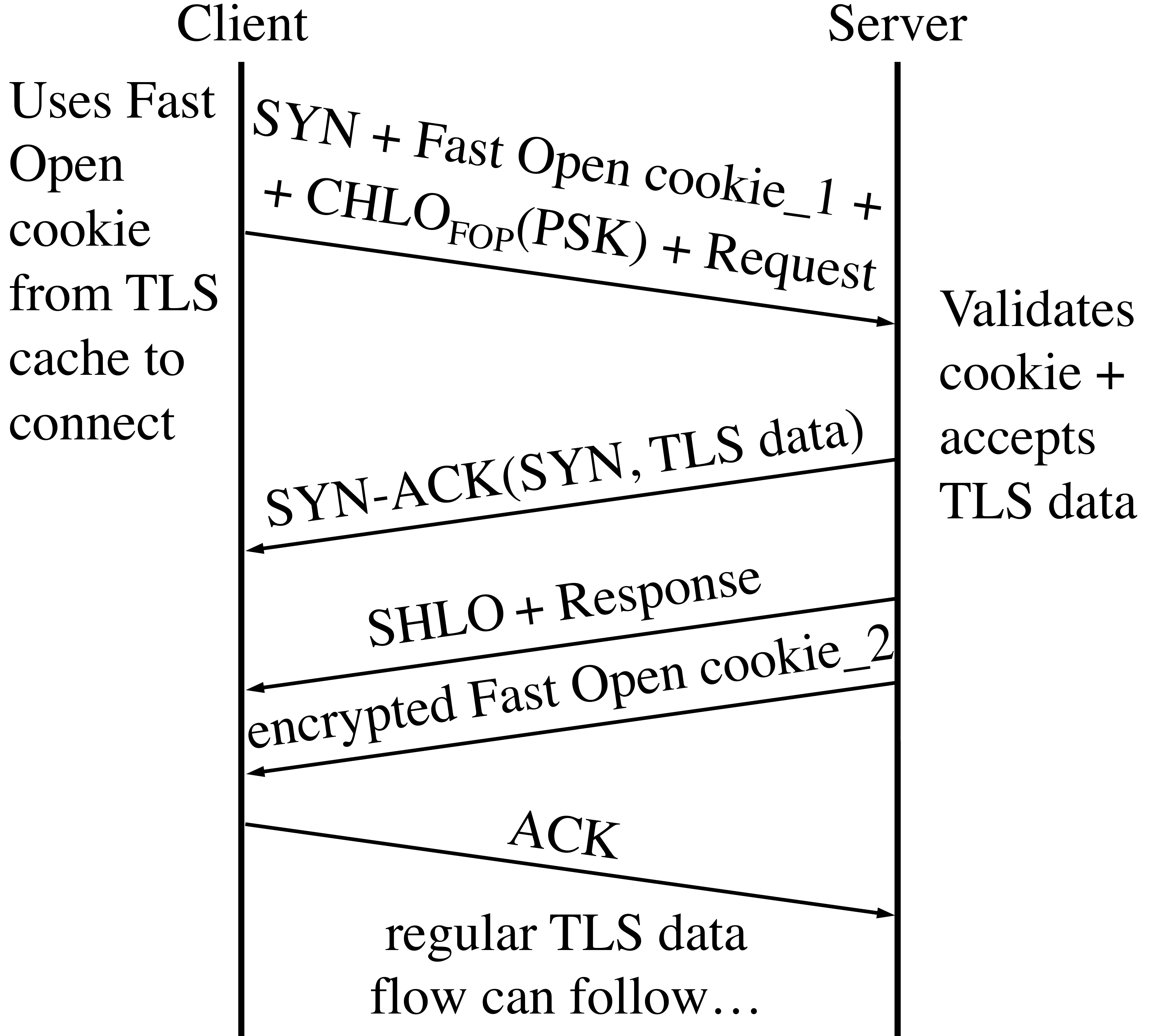}
    \caption*{b) 0-RTT Handshake using TLS Session Resumption}
\end{minipage}
  \caption{Proposed handshakes in TCP Fast Open protocol using TLS 1.3 as a secure channel.}
  \label{fig:proposed_TFO_overview}
  \vskip -12pt
\end{figure*}
\vskip -12pt
\paragraph*{Summary}
As a summary of these browser measurements, we find that the use of the TFO protocol leads to huge privacy risks such as unrestricted tracking periods and cross-browser tracking.
Furthermore, our results indicate that this tracking mechanism is very persistent and cannot be terminated by browser restarts or a change of the browsing mode.
We recommend browser vendors to refrain from deploying the TFO protocol due to the presented privacy problems.

%

\section{TCP Fast Open Privacy}\label{sec:TCP-Fast-Open-Privacy}

In this section, we introduce the TCP Fast Open Privacy (TCP FOP) protocol, that addresses the performance and privacy limitations of TLS over TFO\@.
Then, we describe the implementation of the novel TCP FOP protocol.
Subsequently, we evaluate the privacy and performance provided by the TCP FOP protocol.
Finally, to further substantiate the real-world applicability of TCP FOP, we analyze the effects of TCP protocol entrenchment on the proposed protocol.

\vspace{-1em}
\subsection{Design of TCP FOP} 

TCP FOP builds upon the TFO protocol and continues the approach to use Fast Open cookies to reduce the latency of the connection establishment.
These Fast Open cookies are generated by the server and sent over an encrypted channel to the client.
Because there is no encrypted channel within TCP, we propose to use TLS for this purpose and to create a cross-layer solution.
Our proposal requires an extension to TLS with an additional message type that allows the client to request Fast Open cookies and enables the server to provide such cookies over an encrypted channel. 
The client stores a received cookie along with the corresponding timestamp, the hostname, and a context identifier.
The timestamp is used to limit the period for which a cached cookie can be used to attempt an abbreviated handshake.
The hostname is authenticated within the TLS handshake and can therefore be associated with the cookie.
The cached cookies are then used for abbreviated handshakes for matching hostnames independently of the server's IP address.
This modification allows TCP FOP to anticipate the load balancing of websites across several IP addresses.
However, it requires the involved servers to share the cryptographic secret, that is used to encrypt/decrypt the corresponding Fast Open cookies.
The context identifier is provided by an application and marks the context in which the cookie was retrieved.
Thus, a browser can mark for example each Fast Open cookie from a third-party with the context identifier of the corresponding first party.
As a result, this third-party cannot track the client across several first party websites via the Fast Open cookie because the cached
cookie can only be used for an identical context identifier.
By changing the context identifier for events like browser restarts or changes between browsing modes, user tracking across these events/contexts is mitigated.
Thus, to restrict user tracking via host-based attackers, applications need to provide context identifiers to limit the usage of retrieved Fast Open cookies. 
Additionally, Fast Open cookies should not be reused to set up several abbreviated connections.

In the following, we describe the details of the TCP FOP handshakes.
Figure~\ref{fig:proposed_TFO_overview} shows a schematic of the proposed handshakes that use TLS version 1.3~\cite{rfc8446} as an encrypted channel.

\textbf{Initial handshake:} First, client and server use a standard three-way handshake to establish a TCP connection.
Second, the client starts the modified TLS handshake with a \textit{client hello message} (CHLO$_{\textrm{FOP}}$) that indicates support for TCP FOP\@.
The server responds with a \textit{server hello message} (SHLO) that completes the cryptographic handshake and allows sending subsequent messages over the established TLS channel.
Afterward, the server generates a fresh Fast Open cookie for the client and sends it over the TLS channel.
Finally, the client stores this message along with a timestamp, the hostname, and a context identifier in its TLS cache.

 \begin{table*}[htbp]
   \caption{Kernel API changes}\label{tab:API_changes}
 \centering
 \begin{tabularx}{0.8\linewidth}{lX}
 \toprule
API Name & Description \\
 \midrule
TCP\_Fast\_Open\_COOKIE\_GEN & This server-side API enables an application to retrieve a Fast Open cookie for a specific client from the kernel.\\
TCP\_Fast\_Open\_COOKIE\_SET & This client-side API enables an application to fill the TCP cache with a specific cookie.\\
 \bottomrule
 \end{tabularx}
 \end{table*}

\textbf{0-RTT handshake:} To establish a new secure connection to the same website with an abbreviated 0-RTT handshake, the TLS implementation checks whether a valid Fast Open cookie is available in its TLS cache. 
A Fast Open cookie is valid if it has not yet expired and the respective hostname and context identifier match. 
Assuming a valid cookie, the client then calls a kernel function with the cookie as a parameter that attempts to open a TCP FOP connection.
Then, the client sends a SYN message to the server which includes the respective Fast Open cookie as shown in Figure~\ref{fig:proposed_TFO_overview}b.
Subsequently, the client starts its TLS handshake without waiting for the server's response. Figure~\ref{fig:proposed_TFO_overview}b shows a 0-RTT TLS handshake using session resumption for connection establishment and directly sending an encrypted data request. 
The TLS resumption handshake decreases the delay and saves expensive cryptographic operations compared to a full TLS handshake by leveraging key material exchanged in an earlier  session.
Upon receiving the client's messages, the server validates the contained Fast Open cookie before accepting the payload in the SYN packet. 
The server then accepts the connection and the received TLS data by sending a SYN-ACK message which acknowledges the message of the client.
Otherwise, the server acknowledges the connection only and drops the TLS data, which leads to a rejected 0-RTT handshake.
Following the flow of an accepted 0-RTT handshake attempt as shown in Figure~\ref{fig:proposed_TFO_overview}b, the server's TLS application validates the client's session resumption data.
Assuming this data to be valid, the server answers by sending a SHLO, a response to the client's request and a fresh Fast Open cookie.
In total, this protocol flow (see Figure~\ref{fig:proposed_TFO_overview}b) allows the client to directly send encrypted application data without waiting for a response from the server.

Note, that in TLS version 1.3~\cite{rfc8446} the session resumption mechanism should not reuse identifiers for connection establishment to prevent tracking by a network-based attacker.
To extend this protection against this attacker to the TCP layer, a Fast Open cookie should not be reused for  setting up multiple connections.
\vspace{-1em}
\subsection{Implementation of TCP FOP}

To assess the feasibility of TCP FOP,
we implemented it in the Linux 4.18 kernel and in the wolfSSL TLS library.
Our implementation required only minor modifications
of in total about 300 lines of code (LoC) for Linux and 400 LoC for wolfSSL, including comments and debug output.
\vspace{-1em}
\subsubsection{Kernel support}\label{sec:kernel_patch}

The Linux Kernel 4.18 already supports TCP Fast Open.
Thus, we largely reused the available TCP Fast Open implementation and as a noteworthy change, we added two new APIs to the kernel as shown in Table~\ref{tab:API_changes}. 
The API \textit{TCP\_Fast\_Open\_COOKIE\_GEN} enables the server's TLS application to retrieve a fresh Fast Open cookie for a specific client connection from its kernel.
For the client, we added the inverted API call which allows including a Fast Open cookie into the kernel's cache before the subsequent connection establishment.

Our prototype aims to evaluate that the presented cross-layer approach adds no substantial complexity to TCP and TLS\@.
The implemented \textit{TCP\_Fast\_Open\_COOKIE\_GEN} API works independently of TCP's connection handling.
The \textit{TCP\_Fast\_Open\-\_COOKIE\_SET} API provides a cookie that can be subsequently used within a TCP handshake.
However, this affects only the initialization of TCP's connection establishment. This creates no external constraints on the handshake protocol itself. 
Furthermore, a Linux kernel API to delete specific TCP Fast Open cookies~\cite{tcp_metrics} already exists, which also modifies the initialization of TCP's connection establishment.
The proposed \textit{TCP\_FOP\_COOKIE\_SET} API only provides the logical counterpart to the existing deletion mechanism.
Our implementation of TCP FOP introduces only lightweight modifications to TCP, which lead to no external constraints of TCP's connection handling.

\vspace{-1em}
\subsubsection{TLS support}\label{sec:ssl_patch}

We implemented our prototype as part of the open-source TLS library wolfSSL that provides support for TLS 1.3.
For practicality reasons, we decided to add Fast Open cookies to the session resumption mechanism of TLS 1.3.
However, for the productive use of TCP FOP we recommend an implementation as a dedicated TLS mechanism, i.e., an extension, but leave that for future work.
We use TLS only as a data channel for the cookie and to place the cookie in the TCP cache during the establishment of a new connection.
Therefore, our implementation does not affect TLS's cryptographic components and its connection handling.

In the following, we briefly describe our workaround of implementing TCP FOP by adapting the session resumption mechanism of TLS 1.3.
On the server-side, we include a new Fast Open cookie which we generated with the \textit{TCP\_Fast\_Open\_COOKIE\_GEN} API into each TLS session resumption ticket.
Thus, the TLS server would subsequently send NewTicket messages which contain the Fast Open cookie and the original session resumption ticket.
Upon receiving such a message, the client stores it along with its own connection state such as encryption keys within its TLS cache.
To establish a subsequent connection to the same hostname, the client first validates that a session resumption with that website complies with its privacy configuration.
Assuming that this validation was successful, the client extracts the Fast Open cookie from the cached session resumption ticket and stores it in the kernel's TCP cache using the \textit{TCP\_Fast\_Open\_COOKIE\_SET} API\@.
This API identifies the receiver based on its IP address.
Note, that this approach requires the application to learn the IP address of a given hostname as it is common for applications doing their independent DNS lookups.
In case, the application delegates the name resolution to the kernel, an additional Kernel API would be required to associate a Fast Open cookie with the resolved IP address of a given hostname.
Subsequently, the client uses the session resumption ticket to establish a resumed TLS session over TCP's Fast Open extension.  After this connection is established, the client deletes the used Fast Open cookie within the cache.
To avoid client identification based on session resumption tickets through a network-based attacker, the client shall not reuse the same ticket to set up connections with the server.
Analogous, each fresh session resumption ticket is required to contain a fresh Fast Open cookie, so that a client cannot be identified by linking cookies.
\vspace{-1em}
\subsection{Evaluation of TCP FOP}

 \begin{table*}[htbp]
   \caption{Comparison of privacy characteristics between the TCP Fast Open protocol and our TFO proposal utilizing TLS as a secure channel.}\label{tab:TFO_comparison}
 \centering
 \begin{tabular}{lcc}
 \toprule
\multicolumn{1}{c}{Privacy characteristic} & \multicolumn{1}{c}{TCP Fast Open Protocol} &  \multicolumn{1}{c}{TCP Fast Open Privacy Protocol}\\
 \midrule
Tracking via network-based attacker&viable&blocked through single-use cookies \& encrypted channel\\
Tracking across third-parties&viable&blocking possible through context identifier\\
Tracking across virtual hosts&viable&blocking possible through context identifier\\
Tracking across private browsing modes&viable&blocking possible through context identifier\\
Tracking across browser restarts&viable&blocking possible through context identifier\\
Tracking across user applications&viable&blocking possible through context identifier\\
Tracking across IP address changes&blocked&blocking possible through context identifier\\
Tracking periods&unrestricted&restriction possible through expiration of cookies\\
 \bottomrule
 \end{tabular}
 \end{table*}
 This section starts with an assessment of the privacy properties of TCP FOP and a subsequent comparison to  the previously existing TFO protocol. 
 Next, a performance evaluation of TCP FOP is presented based on experiments with the implemented prototype.
To investigate the real-world applicability of TCP FOP, this section ends with a feasibility analysis studying possible deployment issues. 

 \vspace{-1em}
\subsubsection{Privacy Evaluation}

Tracking via Fast Open cookies is independent of alternative tracking mechanisms such as HTTP cookies, browser fingerprinting, or IP addresses.
To protect the privacy of users, a network-based attacker can observe each utilized Fast Open cookie only once, namely during the 0-RTT handshake of TCP FOP\@.
From the perspective of a network-based attacker, these Fast Open cookies are encrypted and single-use data blocks and thus cannot be linked to a specific user.
Therefore, the TCP FOP protocol prevents attackers to use Fast Open cookies to re-identify specific users and to establish user profiles.
Hence, tracking by network-based attackers is not possible anymore, which is the most important privacy achievement of TCP FOP\@.

However, when facing host-based attackers, the TCP FOP protocol faces a performance versus privacy tradeoff.
The best user privacy is achieved when doing initial handshakes only, while the best performance is achieved during sequences of 0-RTT handshakes.
However, host-based attackers can link 0-RTT handshakes to the same user by linking their Fast Open cookies.
In an initial handshake, the user does not reuse Fast Open cookies from a prior connection, therefore user tracking is prevented.
However, an initial TCP FOP handshake requires an additional round-trip time compared to the 0-RTT connection establishment, which impacts performance. 
TCP FOP provides a mechanism to balance this tradeoff in the context of specific applications.
For that, Fast Open cookies expire after a certain lifetime, which limits the maximum tracking period to this lifetime.
The performance impact of such a lifetime approach has been studied in prior research work~\cite{sy2018tracking}.
This study of users' browsing behavior indicates that 17.7\% of all revisits of websites can use 0-RTT handshakes, if the lifetime of Fast Open cookies is set to five minutes.
Increasing this lifetime to 60 minutes allows using 0-RTT handshakes for 48.3\% of all website revisits.
Thus, this approach allows to strictly enforce an upper limit for the feasible tracking period, while short Fast Open cookie lifetimes still enable a significant share of 0-RTT connection establishments. 

The second countermeasure of the TCP FOP protocol against host-based attackers uses context identifiers associated with cached Fast Open cookies.
These context identifiers are intended to strictly enforce privacy policies in applications and therefore prioritize privacy over performance.
This approach restricts a client to use only cached Fast Open cookies for 0-RTT handshakes if their context identifier is identical to the active context of the application.
Defining the context based on the visited party, virtual host, IP address, browsing mode, user application, and browser session logically excludes the tracking approaches observed in Section~\ref{sec:browser_eval}.
For example, by switching to the private browsing mode, previously cached Fast Open cookies cannot be used for 0-RTT handshakes, as they have been retrieved from the context of a different browsing mode.
However, each additional dependency on the context identifier causes a further restriction for the use of cached Fast Open cookies, which will eventually affect the ratio of initial and 0-RTT handshakes. 
Table~\ref{tab:TFO_comparison} summarizes our findings from Section~\ref{sec:browser_eval} for the TFO protocol and compares them to the privacy characteristics of TCP FOP\@.
We find that the TCP FOP protocol can mitigate all privacy issues of TLS over TFO\@.
\vspace{-1em}
\subsubsection{Performance Evaluation}

 \begin{table*}[tb]
 \caption{Mean duration to establish a connection between the client-server pair and to download a small website using TCP/TLS, TFO/TLS, and TCP FOP/TLS\@. The standard deviation of the mean duration is denoted in brackets. Initial connections use the initial handshake of the respective TCP variant and a full TLS~1.3 handshake. Resumed connections utilize a 0-RTT handshake if supported by the respective TCP variant and a resumed 0-RTT TLS~1.3 connection establishment.}
  \label{tab:TCP_FOP_comparison}
 \centering
 \begin{tabular}{rrrrrrr}
 \toprule
\multicolumn{1}{c}{Network} & \multicolumn{2}{c}{TCP/TLS} & \multicolumn{2}{c}{TFO/TLS} &  \multicolumn{2}{c}{TCP FOP/TLS}  \\
 \cmidrule(){2-3}
 \cmidrule(){4-5}
  \cmidrule(){6-7}
\multicolumn{1}{c}{latency  [ms]} & \multicolumn{1}{c}{Initial [ms]} & \multicolumn{1}{c}{Resumed [ms]} & \multicolumn{1}{c}{Initial [ms]} & \multicolumn{1}{c}{Resumed [ms]}& \multicolumn{1}{c}{Initial [ms]} & \multicolumn{1}{c}{Resumed [ms]}\\
 \midrule
$\approx$0.3 & 28.9 (3.6) & 20.2 (2.7)& 29.9 (3.5)& 22.3 (2.9)& 29.6 (3.6) & 22.2 (2.9)\\
50 ms& 189.8 (2.5)& 132.6 (1.7)&  190.0 (2.4)& 83.7 (1.9)& 190.0 (2.6)& 83.8 (2.2) \\
100 ms & 340.2 (2.1)& 233.1 (1.4)&   340.3 (2.1) & 135.1 (1.6)& 340.7 (2.1) & 135.4 (1.6)\\
150 ms& 490.3 (1.8)& 332.9  (1.3)& 490.7 (1.8)& 185.3 (1.4)& 491.1 (1.8)& 185.7 (1.4)\\
 \bottomrule
 \end{tabular}
 \end{table*}

We evaluate the performance of the TCP FOP protocol in two parts:
First, we conduct experiments to investigate whether the usage of the proposed TCP FOP/TLS incurs a delay overhead compared to connections using TFO/TLS or standard TCP/TLS\@.
Second, we study the performance of TCP/TLS, TFO/TLS, and TCP FOP/TLS in a scenario with real-world load-balancing.
\vspace{-1em}
\paragraph{Experiment using the TCP FOP Prototype}
We compare our implemented prototype of TCP FOP to implementations of standard TCP and TFO\@.
For that, we compare the required time to download a small web page from a single host using one of these three transport protocols in combination with TLS 1.3.
For this experiment, we use two virtual machines, one acting as web server and the other as client.
The virtualization is realized on the same host using  qemu 2.8 and libvirt 3.0.0.
This test setup leads to short network latencies with an average ping of 0.3 milliseconds (ms) between the virtual machines.
The host of the virtual machines was equipped with an Intel Xeon E5-1660 v4 CPU with 32~GB of RAM and ran Debian stretch.
The client and the server machines were set up with 4~GB of RAM and were running 
an Ubuntu 18.10 with our modified Linux kernel (see Section~\ref{sec:kernel_patch}) that supports the TCP FOP protocol.
The server ran the example server program shipped with our modified wolfSSL library.
The program responds to a successful connection establishment with a short string.
The client ran the corresponding example client program of wolfSSL that establishes a TLS session to the server, waits for the short string from the server and terminates the TLS session upon its reception.
Note that we used our modified wolfSSL implementation as described in Section~\ref{sec:ssl_patch} for this measurement.
In our experiment, we established and resumed a new TLS connection via standard TCP, TFO, or TCP FOP\@.
All of these TLS connections used the forward-secure cipher suite \textit{TLS\_AES\_128\_GCM\_SHA256}.
The initial TLS connection uses an initial handshake of the corresponding TCP variant.
The 0-RTT TLS resumption handshake uses an abbreviated TCP connection establishment if supported by the respective TCP variant.
To account for skews in the measurements, we repeated the experiment \num{1000} times and measured the elapsed wall-clock time.
We conducted our measurements with the client's network interface configured to simulate network latencies of 0.3~ms, 50~ms, 100~ms, and 150~ms with iproute2's \texttt{tc} program.
We recorded and inspected the network traffic of the virtual network interface to validate a correct behavior of our evaluation setup.

The results of the measurement are shown in Table~\ref{tab:TCP_FOP_comparison}.
We find that for minimal network latencies of 0.3~ms TLS over the standard TCP provides the best performance results, while the usage of TFO/TLS and TCP FOP/TLS leads to similar values.
Especially, in the case of a resumed handshake the usage of the standard TCP provides a performance gain of almost ten percent.
We assume that TFO and TCP FOP have a computational overhead by generating, validating, and handling the Fast Open cookies that incurs this delay.
 
In total, our results indicate only small differences between the performance of the initial handshake measurements that are consistently less than a millisecond between the investigated TCP variants for the same network latency.
For resumed handshakes, the performance benefits of TFO/TLS and TCP FOP/TLS are significant for larger network latencies.
These protocols complete the resumption handshake with time savings larger than 50\% for network latencies of 50~ms and above compared to TLS over standard TCP\@.
These benefits account for the saved round-trip time of the 0-RTT handshakes of TFO and TCP FOP\@.
Between the performance of TFO/TLS and TCP FOP/TLS, we find only insignificant differences.
As a result of this measurement, we find that TFO/TLS and TCP FOP/TLS have similar computational overhead.  
 
\vspace{-1em}
\paragraph{Simulation considering Load-balancing}
Clients using the TCP FOP/TLS associate a retrieved Fast Open cookie with the hostname of the respective online service. This allows them to attempt 0-RTT handshakes with online services with matching hostnames independently of the server's IP address. Conversely, the TFO/TLS is restricted to conduct 0-RTT handshakes only when the server's IP address matches the one associated with the cached Fast Open cookie. This simulation investigates the performance benefits of this adapted design of the TCP FOP/TLS protocol stack.

Our test setup consists of a client, a network link, and a website.
For the network link, we assume a mobile LTE connection, with a round-trip time of 60~ms as it is common in the U.S.~\cite{opensignal}.
Note, that the round-trip time for 3G and WiFi connections are on average longer than for LTE connections~\cite{opensignal2}.
Statistically, an average website requires 20 TCP/TLS connections to several hosts for its retrieval~\cite{sy2019enhanced}. To resemble a real-world website, our test website directly links to 19 resources, each of them on a separate host. From the perspective of a domain tree, these 19 hosts are on the same hierarchical level. Furthermore, we assume that all hosts in this test setup support the TCP FOP protocol and use on average the same load-balancing approaches as observed in Section~\ref{sec:performance} for hosts supporting TFO\@.
The client measures the elapsed wall-clock time to establish connections to all 20 hosts that need to be involved to retrieve the website.

 \begin{table*}[htbp]
   \caption{Analysis of the delay overhead of TFO/TLS and TCP FOP/TLS compared to TCP/TLS\@. We simulate the retrieval of a sample website and consider load-balancing as observed in Section~\ref{sec:performance}. We assume a RTT of 60ms for the LTE connection.}\label{tab:PerformanceTCPFOPsimulation}
 \centering
 \begin{tabular}{lrrrrrr}
\toprule
  \multicolumn{1}{c}{} & \multicolumn{2}{c}{$1^{st}$ Revisit} & \multicolumn{2}{c}{$2^{nd}$ Revisit} &  \multicolumn{2}{c}{$3^{th}$ Revisit}  \\
 \cmidrule(){2-3}
 \cmidrule(){4-5}
  \cmidrule(){6-7}
\multicolumn{1}{c}{Simulation}& \multicolumn{1}{c}{TFO/TLS} & \multicolumn{1}{c}{TCP FOP/TLS} & \multicolumn{1}{c}{TFO/TLS} & \multicolumn{1}{c}{TCP FOP/TLS} & \multicolumn{1}{c}{TFO/TLS} & \multicolumn{1}{c}{TCP FOP/TLS}\\
 \midrule
 Probability to save zero RTT& 39.3\% & 0.0\% & 24.6\% & 0.0\% & 8.1\% & 0.0\%\\
 Probability to save one RTT& 60.7\% &  0.0\% & 75.1\% & 0.0\% & 78.5\% & 0.0\%\\
 Probability to save two RTT & 0.0\% & 100.0\% & 0.3\% & 100.0\% & 13.4\% & 100.0\%\\
Mean delay overhead over LTE & -36.4 ms & -120.0 ms & -45.5 ms & -120.0 ms & -63.1 ms & -120.0 ms\\
\bottomrule
 \end{tabular}
 \end{table*}

Table~\ref{tab:PerformanceTCPFOPsimulation} summarizes the results for successive revisits of the test website.
As the TCP FOP/TLS protocol stack can establish 0-RTT connections independently of the IP address associated with a hostname, it saves on each revisit of the website two round-trip times compared to the initial website visit.
One RTT can be saved when connecting to the primary host, and another RTT can be saved by successfully establishing 0-RTT connections to the 19 secondary hosts.
For each website revisit with the TCP FOP/TLS, this reduces the delay overhead for establishing connections to all 20 hosts by 120~ms compared to the initial website visit.
As indicated in Figure~\ref{fig:failure_rate}, the failure rate of the TFO/TLS protocol stack depends on the number of prior visits to a website.
We used a tree diagram to compute the probabilities of saving zero, one, or two RTT during the connection establishment with all 20 hosts.
We find, that for the first revisit the probability of saving a RTT during the connection establishment to all hosts is 60.7\% and on average, the delay overhead is reduced by 36.4~ms.
Note, that the saving of the TCP FOP/TLS protocol stack for the same task is more than three times higher with 120~ms.
For the second and third revisit to the website, the achieved reductions are 45.5~ms and 63.1~ms, respectively. Thus, we observe that the TCP FOP/TLS protocol stack significantly outperforms TFO/TLS, if real-world load-balancing of websites is considered.

\vspace{-1em}
\subsubsection{Feasibility Analysis}

Efforts to deploy alternative TCP versions on the Internet indicate that middleboxes such as Firewalls and NAT devices modify or block unfamiliar TCP packets~\cite{raiciu2012hard}.
Hence, even simple changes to TCP require a long time before they exhibit a significant deployment~\cite{honda2011still}.
The TFO protocol faces similar problems that hinder its rapid deployment on the web~\cite{tfo_deploy, rfc7413}.
The presented TCP FOP embeds a standard TCP three-way handshake in the initial handshake.
This standard handshake is common on the web and thus will not cause any issues with middleboxes~\cite{raiciu2012hard}.
The exchanged messages in TCP FOP's 0-RTT handshake are identical to the messages in the TFO protocol (see Figure~\ref{fig:TFO_overview}b and c).
Note, that for privacy reasons the \textit{cookie\_2} of a rejected 0-RTT handshake should be discarded by the client and not be used to establish new TCP FOP connections.
These identical protocol flows for TFO's and TCP FOP's 0-RTT handshakes avoid further deployment issues.
Thus, middleboxes supporting TFO's 0-RTT handshake will by default also support the TCP FOP protocol.
As a result, the deployment of TCP FOP/TLS on the Internet is not causing additional compatibility issues beyond the ones from the TFO/TLS protocol stack.

Compared to the usage of TFO, the cross-layer optimization TCP FOP requires TLS libraries to control Fast Open cookies.
As a drawback, this interconnects TCP and TLS impacting the portability and maintenance of TCP FOP\@.
However, with \textit{Kernel TLS} there exists an implemented example for a performance-optimization causing a similar drawback between TLS libraries and Kernel functions~\cite{K_TLS}.
In detail, \textit{Kernel TLS} aims to save resources by delegating some expensive computations from the user space to the kernel space.
Overall, we find that TCP FOP provides sufficient performance and privacy benefits to justify a cross-layer solution.

\vspace{-1em}
\paragraph*{Summary}
Based on our evaluation, we find that TCP FOP/TLS significantly improves the privacy and real-word performance of TLS over TFO\@. With respect to privacy, all identified privacy issues of TLS over TFO can be addressed by TCP FOP/TLS\@.
From a performance perspective, the TCP FOP/TLS protocol stack outperformed TLS over TFO in our test scenario that resembles the retrieval of an average real-world website.
Furthermore, we conclude that TCP FOP/TLS will experience fewer deployment issues than TLS over the TFO protocol.
\vspace{-1em}
\section{Related Work}\label{sec:Related}

To the best of our knowledge, we are the first to report on the privacy aspects of the TFO protocol.
While research work on privacy issues within TCP that allows distinguishing clients or servers based on their TCP timestamps~\cite{murdoch2006hot, polcak2014comment} exists,
our reported storage-based tracking mechanism is unrelated to such tracking approaches via the TCP timestamps.
A similar tracking mechanism has been reported for the QUIC transport protocol~\cite{sy2019quic}. 
However, QUIC's address validation tokens are distributed via an encrypted channel and do not allow a passive network observer to link different connections to the same user.
Furthermore, TLS session resumption mechanisms~\cite{sy2018tracking} enable user tracking.
Compared to the TFO protocol, Session resumption in TLS~1.3 does not enable a passive network observer to track a user's online activities.
Thus, the TFO protocol provides substantially lower privacy guarantees than TLS~1.3.
Additionally, prior research includes a proposed TCP extension which directly allows encrypting TCP packets~\cite{bittau2010case}.
This approach significantly increases the complexity of TCP by including typical TLS functionality.
However, our proposed TCP FOP aims to be a lightweight protocol modification where Fast Open cookies are issued via a TLS-encrypted channel.
The limitation of the TFO protocol to anticipate load balancing with multiple server IP addresses has been pointed out in prior work~\cite{langley2017quic}.
We contributed by investigating this limitation under real-world conditions and find that approx.~40\% of the first website revisits fail to establish an abbreviated connection setup. 
TCP FOP presents a novel protocol which anticipates web server load balancing to achieve a higher share of abbreviated handshakes and fully protects against tracking via network-based attackers.
To the best of our knowledge, we are the first to present such a cross-layer approach for the TLS over TFO stack.
\vspace{-1em}
\section{Conclusion} \label{sec:Conclusion}

TFO provides considerable latency improvements compared to TCP's three-way handshake.
However, its usage on the Internet raises alarming privacy concerns.
Therefore, we urge vendors of operating systems and browsers to discourage the deployment of the TFO protocol.
To address the privacy problems at hand, we designed and implemented the TCP FOP protocol.
Our analysis indicates that  TLS over TCP FOP protects against user tracking by network-based attackers.
Furthermore, the protocol can also restrict third-party tracking, enables each application to control its privacy properties and to balance the trade-off between lower delays and privacy protection.
To that end, applications can influence the lifetime of cached Fast Open cookies, which represents the maximum feasible tracking period.
TCP FOP/TLS not only provides better privacy protection, but it also provides significant performance gains in terms of delay compared to TLS over TFO\@.
TCP FOP/TLS can carry out abbreviated TCP handshakes even when the website is served from multiple IP addresses, e.g., when being part of server load balancing.
Our measurements indicate that TFO/TLS fails to establish an abbreviated handshake with a chance of 39.3\% during the first revisit of a website.
We attribute this mainly to server load balancing under which TCP FOP/TLS will always be able to carry out an abbreviated handshake.
We conclude that based on our evaluation the proposed TCP FOP protocol leads to substantial enhancements of the performance and privacy of TLS over TCP Fast Open.

\section*{Acknowledgment}
We thank the anonymous reviewers for their insightful comments and suggestions.
We also thank our shepherd, Alan Mislove, for providing valuable feedback and guidance in the revision process.
This work is supported in part by the German Federal Ministry of Education and Research under the reference numbers 16KIS0381K and 16KIS0922K.

\bibliographystyle{ACM-Reference-Format}
\bibliography{sample-bibliography}

\end{document}